\documentclass[aps,prl,twocolumn,superscriptaddress,showpacs,floatfix]{revtex4-1}
\usepackage{graphicx}
\usepackage{color}
\begin{document}

\title{Experimental demonstration of a Hadamard gate for coherent state qubits}
\author{Anders Tipsmark}
\email[Corresponding author: ]{anders.tipsmark@fysik.dtu.dk}
% \email[Corresponding author: ]{anders.tipsmark@gmail.com}
\affiliation{Department of Physics, Technical University of Denmark, Fysikvej, 2800 Kgs.~Lyngby, Denmark}
\author{Ruifang Dong}
\affiliation{Quantum Frequency Standards Division, National Time Service Center (NTSC), Chinese Academy of Sciences, 710600 Lintong, Shaanxi, China}
\affiliation{Department of Physics, Technical University of Denmark, Fysikvej, 2800 Kgs.~Lyngby, Denmark}
\author{Amine Laghaout}
\affiliation{Department of Physics, Technical University of Denmark, Fysikvej, 2800 Kgs.~Lyngby, Denmark}
\author{Petr Marek}
\affiliation{Department of Optics, Palack\'{y} University, 17.~listopadu 12, 77146 Olomouc, Czech Republic}
\author{Miroslav Je\v{z}ek}
\affiliation{Department of Optics, Palack\'{y} University, 17.~listopadu 12, 77146 Olomouc, Czech Republic}
\affiliation{Department of Physics, Technical University of Denmark, Fysikvej, 2800 Kgs.~Lyngby, Denmark}
\author{Ulrik L. Andersen}
\affiliation{Department of Physics, Technical University of Denmark, Fysikvej, 2800 Kgs.~Lyngby, Denmark}

\date{\today}

\begin{abstract}
We discuss and experimentally demonstrate a probabilistic
Hadamard gate for coherent state qubits. The scheme is based
on linear optical components, non-classical resources and
the joint projective action of a photon counter and a homodyne
detector. We experimentally characterize the gate for the
coherent states of the computational basis by full tomographic
reconstruction of the transformed output states. Based on the
parameters of the experiment we simulate the fidelity for all
coherent state qubits on the Bloch sphere.
\end{abstract}

\pacs{03.67.-a,03.67.Lx,42.50.Ex}
% 03.67.-a  Quantum information
% 03.67.Lx	Quantum computation
% 42.50.Ex  Optical implementations of quantum information processing
%\keywords{}

\maketitle

Measurement-based, linear optical quantum processors rely on
off-line prepared resources, linear optical transformations
and measurement-induced operations \cite{RevModPhys.79.135}.
Among all measurement-based protocols, the most famous ones
are the cluster state quantum processor where universal
operations are executed by measuring a large entangled
cluster state \cite{PhysRevA.68.022312},
and the linear quantum computer approach proposed by Knill,
Laflamme, and Milburn \cite{Nature.409.6816.46}. The latter
method is based on single photon resources that interfere
in a linear optical network and subsequently are measured
to enforce the desired operation. Despite its seeming
simplicity, the implementation of a fault tolerant operating
algorithm is complex as it requires a very large overhead. 

An alternative approach to measurement based linear quantum
computing has been put forward by Ralph \emph{et al.}
\cite{ISP7784948}. Rather than using discrete degrees of
freedom (e.g. the polarization) of a single photon as the
computational basis, it was suggested to use two mesoscopic
coherent states, $|\alpha\rangle$ and $|\!-\!\alpha\rangle$, where
$\alpha$ is the amplitude. Although these states are only
approximately orthogonal ($\langle\alpha|-\alpha\rangle \neq 0$),
resource efficient and fault tolerant quantum gates can be
implemented: For a large coherent amplitude,
that is $\alpha >2$, deterministic gates can in
principle be realized although the experimental
implementation is very challenging \cite{QuantumInformation}.
On the other hand, by employing a simpler physical implementation,
non-deterministic gates can be realized for any value of $\alpha$,
and for $\alpha>1.2$, the scheme was theoretically shown to be
fault-tolerant and resource efficient \cite{PhysRevLett.100.030503}. 

An even simpler implementation of a universal set of
non-deterministic quantum gates was recently suggested
by Marek and Fiur\'{a}{\v{s}}ek \cite{PhysRevA.82.014304}.
They proposed the physical realization of a single mode
and a two-mode phase gate as well as the Hadamard gate.
In this Letter we present a proof of principle experiment
of the probabilistic Hadamard gate for coherent state
qubits. The implemented protocol is based on
a squeezed state resource, linear operations as well as
two projective measurements of discrete and continuous
variables. By injecting the computational basis states
($|\alpha\rangle$ and $|-\alpha\rangle$) into the gate
we characterize its function by reconstructing the Wigner
functions of the transformed output states and calculate
the fidelity with ideally transformed state.
The implementation of the Hadamard gate demonstrated
in this Letter constitutes the very first step towards
the realization of a quantum processor based on coherent
state qubits.

\begin{figure}[!b]
		\includegraphics[width=1.0\columnwidth]{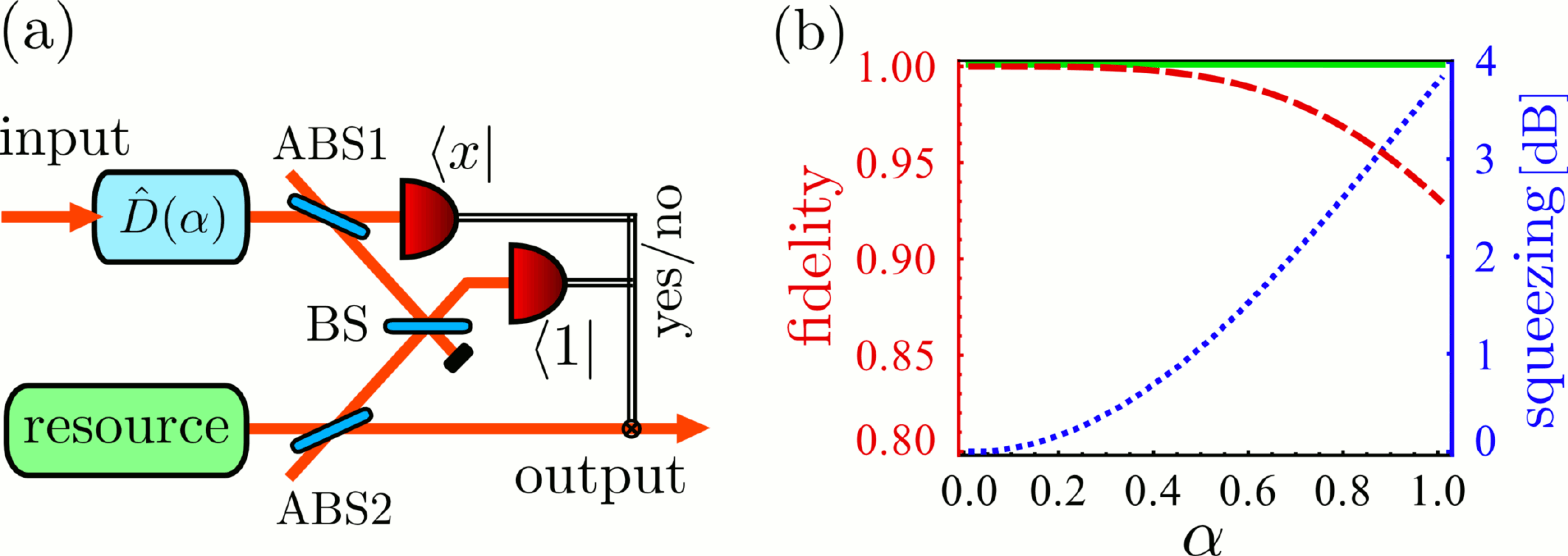}
	\caption{(Color online)
(a) Schematic of the Hadamard gate.
The input coherent staten qubit (CSQ) is displaced ($\hat{D}$)
and mixed with a resource state at a beam splitter (BS). The output
of the gate is conditioned by a single photon detection ($\langle 1|$)
and a homodyne measurement ($\langle x|$).
(b) Gate fidelity as a function of the CSQ amplitude for an ideal
coherent state superposition resource (solid green) and the
squeezed state resource (dashed red). The degree of squeezing
that optimizes the fidelity is represented by the dotted blue curve.}
	\label{fig:Gate_Schematic}
\end{figure}

A Hadamard gate transforms the computational basis states,
$|\pm\alpha\rangle$, into the diagonal basis states,
$(|\alpha\rangle\pm|\!-\!\alpha\rangle)/\sqrt{N_{\pm}}$,
which we refer to as the even and odd coherent state qubits
(CSQ) \cite{csq_0,csq_1,csq_2,csq_3,csq_4,csq_5,csq_6,note_1}.
Such a transformation can be performed probabilistically
using the circuit shown in Fig.~\ref{fig:Gate_Schematic}~(a).
The gate is based on a supply of coherent state superposition
resources which are assumed to have the same amplitude as the
coherent states of the computational basis. The gate works by displacing the
arbitrary CSQ input state $|\psi_{\rm in}\rangle=
(u|\alpha\rangle+v|\!-\!\alpha\rangle)/\sqrt{N}$
followed by a non-distinguishable subtraction of a single photon, from either the displaced input or the resource state. Physically, this can be done by reflecting a small part of either state
using highly asymmetric beam splitters (ABS1,ABS2), interfering
the resulting beams on a beam splitter (BS) with transmittivity
$t$ and reflectivity $r$, and detecting one photon at the
output with a single-photon detector. Theoretically this
is described by the operator $r\hat{a}+t\hat{b}$ where $\hat{a}$
and $\hat{b}$ are annihiliation operators corresponding
to the subtraction of a photon from the displaced input and
the coherent state superposition resource, respectively.
As a final step the two-mode
state is projected onto the single-mode quadrature eigenstate
$|x\rangle$, where $x$ is the amplitude quadrature, by using
a homodyne detector (HD). The resulting output state is
\begin{eqnarray}
  u\frac{|\alpha\rangle+|-\alpha\rangle}{\sqrt{N_+}}+
  Y_1(u+vZ)\frac{|\alpha\rangle-|-\alpha\rangle}{\sqrt{N_-}},
\end{eqnarray}
where
\begin{eqnarray}
  Y_1~=~\frac{t}{2r}\sqrt{\frac{N_-}{N_+}},\qquad
  Z~=~\frac{\langle x|0\rangle}{\langle x|2\alpha\rangle}.
\end{eqnarray}
By using a beam splitter (BS) with $t\ll r$ and setting the $x$
quadrature such that $Z\gg 1$ and $Z Y_1=1$, the Hadamard transform is
implemented. The gate is probabilistic, and implemented by a hybrid
detection system, using both discrete and continous variable projections
\cite{hybrid_1,hybrid_2}. Its success is conditioned on the joint
measurement of a photon and a quadrature measurement
outcome with the value $x$.

As an even coherent state superposition with small amplitude
is reminiscent of a squeezed vacuum state, and this latter state
is experimentally easier to prepare, we will in the following
consider the replacement of the ideal resource with a squeezed
vacuum state. With this substitution,
the transformed state will have the following form,
\begin{eqnarray}
  u\hat{S}(s)|0\rangle+Y_2(u+vZ)\hat{S}(s)\hat{a}^\dagger|0\rangle,
\end{eqnarray}
where $s$ is the squeezing parameter which is related to the
squeezing variance by $V=e^{-2s}$, and the parameter $Y_2$ is now
given by
\begin{eqnarray}
%  Y_2~=~-\frac{t\sinh(s)}{2r\alpha}.
  Y_2 = -t\sinh(s)/(2r\alpha).
\end{eqnarray}
Again, the requirement for optimal implementation of the Hadamard
transform is $Z\gg 1$ and $Z Y_2=1$.
Using this result we calculate the expected gate fidelity for
various amplitudes $\alpha$ as shown by the dashed red curve in
Fig.~\ref{fig:Gate_Schematic} (b).
For the squeezed vacuum resource, we optimize the squeezing degree
(shown by the dotted blue curve) to obtain the highest fidelity
which reaches unity for $\alpha=0$. At higher amplitudes,
the resource deviates from the ideal coherent state superposition
and thus the fidelity decreases.
For comparison, we also plot the expected gate fidelity for the
case of an ideal resource (the solid green line).
In the experiment described below we use $\alpha=0.8$
which gives a reasonable trade-off between fidelity ($F=0.97$),
required squeezing ($V=2.6$~dB) and success probability.

\begin{figure}[!b]
	\centering
		\includegraphics[width=0.9\columnwidth]{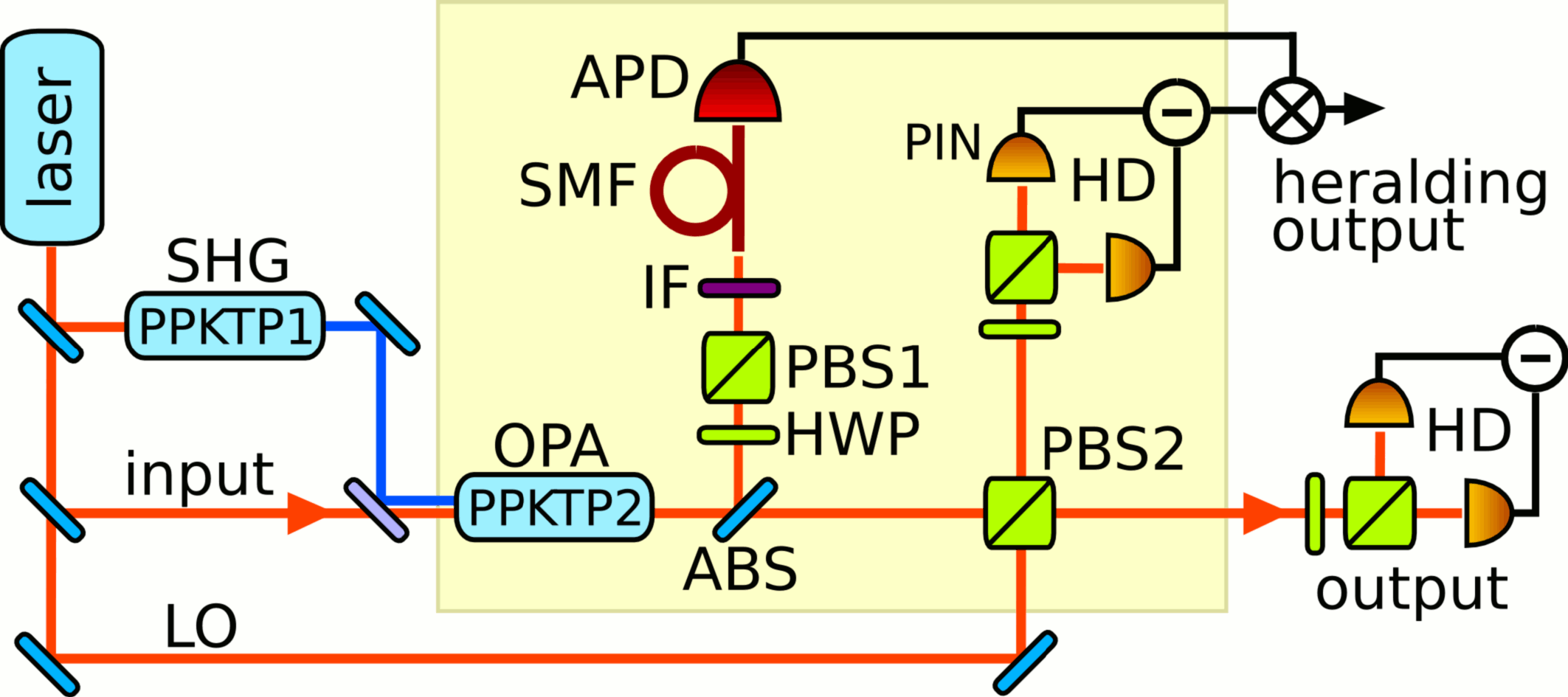}
	\caption{(Color online)
	Experimental setup for the coherent state qubit Hadamard gate.
% 	The up-converted (SHG) pulsed laser is used for the preparation of
% 	a squeezed state resource (OPA) and as a coherent input of the gate.
% 	The small portion of the laser serves as a local oscillator (LO).
%   The yellow-highlighted box denotes the core part of the Hadamard gate
%   with a classical heralding output and the CSQ output which is fully
%   characterized by means of the homodyne detection (HD).
  }
	\label{fig:Setup_Hadamard}
\end{figure}

The experimental setup is presented in Fig.~\ref{fig:Setup_Hadamard}.
Nearly Fourier-limited picosecond pulses ($4.6$~ps) generated
by a cavity dumped Ti:sapphire laser with repetition rate of
$815$~kHz and central wavelength of $830$~nm are frequency doubled
(SHG) by single passing a $3$~mm long periodically poled KTiOPO$_4$
nonlinear crystal (PPKTP1). Up-converted pulses at $415$~nm pumps
a second crystal (PPKTP2) which is phase-matched for degenerate
collinear optical parametric amplification (OPA), thus
yielding up to $3$~dB of vacuum squeezing, in the vertical polarization.
This state is used as a resource for the Hadamard gate.
An adjustable fraction of a horizontally polarized mode
at $830$~nm passes the OPA crystal unchanged and serves
as the input coherent state to the gate.
About $7.5\%$ and $1.5\%$ of the coaxially propagating resource
and input modes, respectively, are reflected off an asymmetric
beam splitter (ABS) and transmitted through a half wave plate
(HWP) and a polarizing beam splitter (PBS1) which in combination
acts as a variable beam splitter (BS) thus mixing the input mode
and the resource mode. The transmittance $|t|^2$ of the BS is set
to $25\%$. %% \cite{note2}.
The output is spatially and spectrally
filtered by a single mode optical fiber (SMF) and a narrow interference
filter (IF) with a bandwidth of $0.05$~nm and detected by
a single photon counting module based on a silicon avalanche
photo diode (APD) with dark count rate of $20\pm 4$ per second.
The total efficiency of the APD arm reaches $25\pm 4\%$.

The transmitted fraction of the modes after the asymmetic beam
splitter is superimposed with a bright local oscillator (LO)
at a polarizing beam splitter (PBS2).
The amplitude quadrature is measured on the reflected mode
by homodyne detection with a fixed relative phase set to zero.
The recording of the measurement results was done by correlating
the APD detection events with a synchronization signal from the
laser cavity dumper through a coincidence circuit to decrease
the probability of dark events. Every time a photon
was detected by the APD within the accepted time slot,
the homodyne signal was sampled by an oscilloscope
running in memory segmentation regime and fed to a computer
where the corresponding quadrature value was processed.
The state at the output of the gate is
measured with another homodyne detector with the relative
phase of the LO scanned over a period and then
reconstructed using maximum-likelihood based quantum state
tomography \cite{MaxLik}. In the reconstruction we corrected for the total detection efficiency of the homodyne detector, which was estimated to be $77\pm2\%$, including efficiency of the photo diodes, ($93\pm1\%$), visibility, ($95\pm1\%$) and transmission efficiency, ($93\pm1\%$).

Making a full experimental investigation of the gate performance would require access to arbitrary coherent state qubits, $u|\alpha\rangle+v|-\alpha\rangle$. These states are themselves experimentally challenging to prepare and  experiments covering the entire bloch sphere have only reached fidelities of $50-70\%$ \cite{csq_0}. The quality with which we could expect to prepare these states would in itself compromise the applicability in characterizing the gate performance. For this reason we have chosen to only investigate the performance of the Hadamard gate with
%In our experiment, we investigated the performance of the
%Hadamard gate for 
the computational basis states $|\pm\alpha\rangle$ as inputs, since these states can be prepared with unit fidelity. After the displacement operation, $\hat{D}(\alpha)$,
this corresponds to the injection of $|0\rangle$ and
$|2\alpha\rangle$, where $\alpha=0.8\pm0.2$ in our case.
The uncertainty is due to the imperfect calibration
of total losses of the whole setup. As described, the gate
is heralded by conditioning on two different measurement
outcomes---the APD detection event and a certain outcome
of the first homodyne detector. It can be seen that
the conditional homodyning only plays a role when
we inject a CSQ into the gate,  i.e. when ${u,v}\neq0$.
With coherent states as the input, the solution is to choose
a narrow heralding window that would balance
the success probabilities of the gate for those basis states.
For the input state $|\!-\!\alpha\rangle$ the APD detection
probability was of the order of $10^{-3}$ while for the
$|\alpha\rangle$ input state, the probability was of the
order of $10^{-2}$. From this we can see that we need to
choose a heralding window that will balance out the factor
of $10$. Based on the experimental data we found its optimal
position $x=0.4$
% x = ( eta Ttap 4 alpha^2 + ln( sqrt(P0/Pbeta)))/ (4 sqrt( eta Ttap) alpha)
and the width of $0.02$ that would give us an overall
success probability of the order of $10^{-5}$.

\begin{figure}[b!]
\includegraphics[width=0.85\columnwidth]{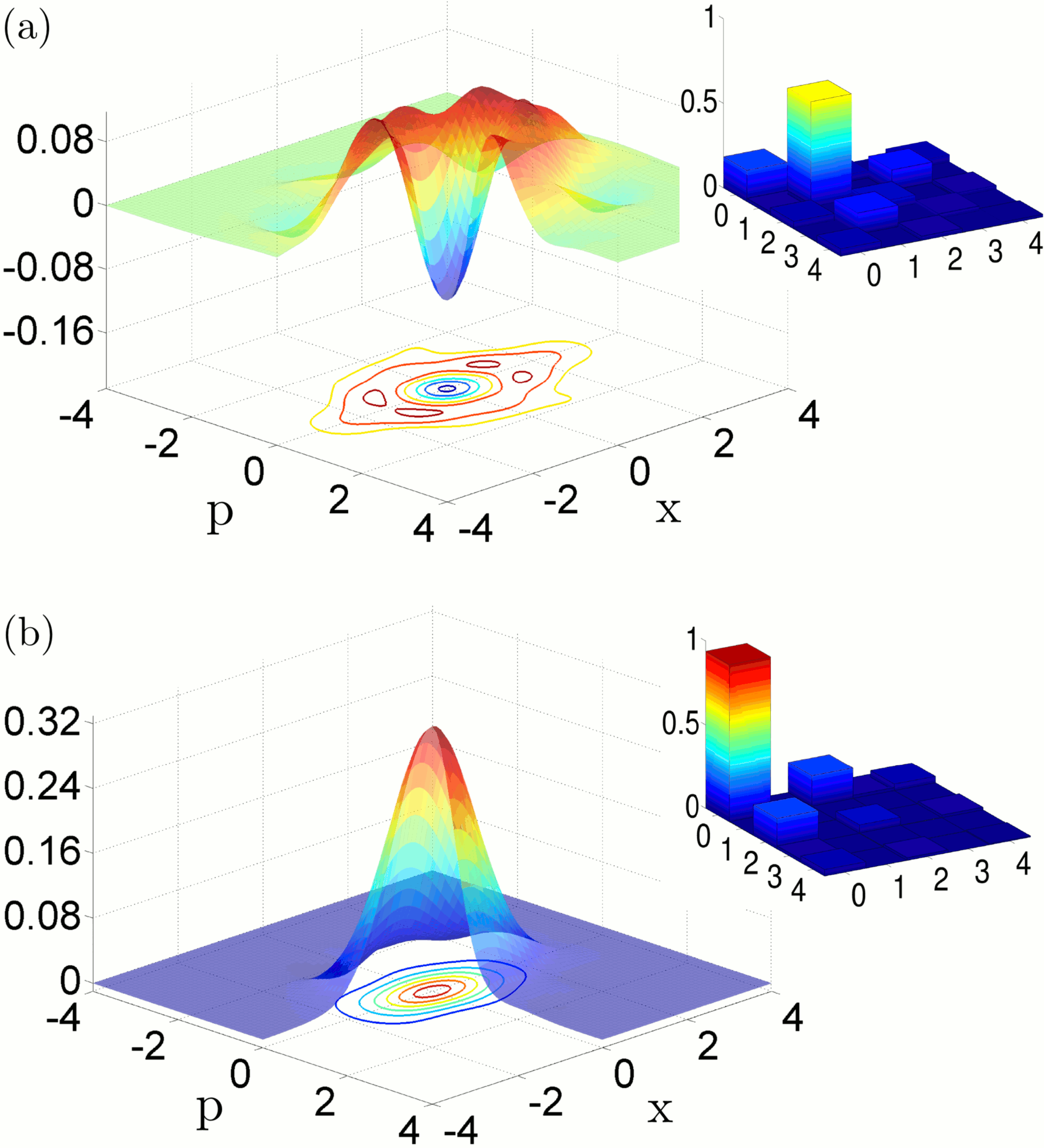}
  \caption{Reconstructed density matrices (insets) and calculated Wigner
  functions of the output states for (a) $|\!-\!\alpha\rangle$ input
  and (b) $|\alpha\rangle$ input.}
	\label{fig:Results}
\end{figure}

The reconstructed output states for both input states
$|\!-\!\alpha\rangle$ and $|\alpha\rangle$ can be seen
in Fig.~\ref{fig:Results}.
For the $|\!-\!\alpha\rangle$ input, the gate yields a state
which closely resembles a small odd cat state which is what
we expect from the gate operation. We found the fidelity
between the prepared state and the ideal CSQ,
$(|\alpha\rangle-|\!-\!\alpha\rangle)/\sqrt{N_{-}}$, is maximized
for $\alpha=0.75$ and reaches the value of $F_{-\alpha}=0.65\pm0.04$.
The non classicality of the superposition state produced
by the Hadamard gate can be seen from the negativity
of the corresponding Wigner function which is
$W(0,0)=-0.11\pm0.02$, which is comparable to previous experiments where photon subtraction has been used to prepare non-Gaussian states \cite{csq_0,csq_1,csq_2,csq_3,csq_4,csq_5,csq_6}. The non-classical effects was also observable without correction, with a fidelity of $F_{-\alpha}=0.55\pm0.04$ and a value at the origin of $W(0,0)=-0.05\pm0.02$.
%{\color{blue} I do not know if we should mention results before correction. $F_{-\alpha}=0.55\pm0.04$ and $W(0,0)=-0.05\pm0.02$ approximately.}
For the $|\alpha\rangle$ input, the output state closely resembles
a squeezed state, aproximating a small even CSQ,
$(|\alpha\rangle+|\!-\!\alpha\rangle)/\sqrt{N_{+}}$. The fidelity between
the prepared state and the ideal CSQ for $\alpha=0.75$
was found to be $F_{\alpha}=0.94\pm0.02$.

The experimental results shown in Fig.~\ref{fig:Results} only provide
a partial test of the Hadamard gate. In order to evaluate its action
on an arbitrary CSQ as the input, we conducted a numerical simulation
of the gate taking into account all the important experimental
imperfections, namely the realistic splitting ratios of ABS1, ABS2 and BS,
losses in APD and HD channels, and the impurity of our resource
squeezed state. The effect of APD false clicks
%% due to electronic dark counts and non-unity spatial and spectral purities
was included as well, but found to be negligible.

\begin{figure}[t!]
\includegraphics[width=0.9\columnwidth]{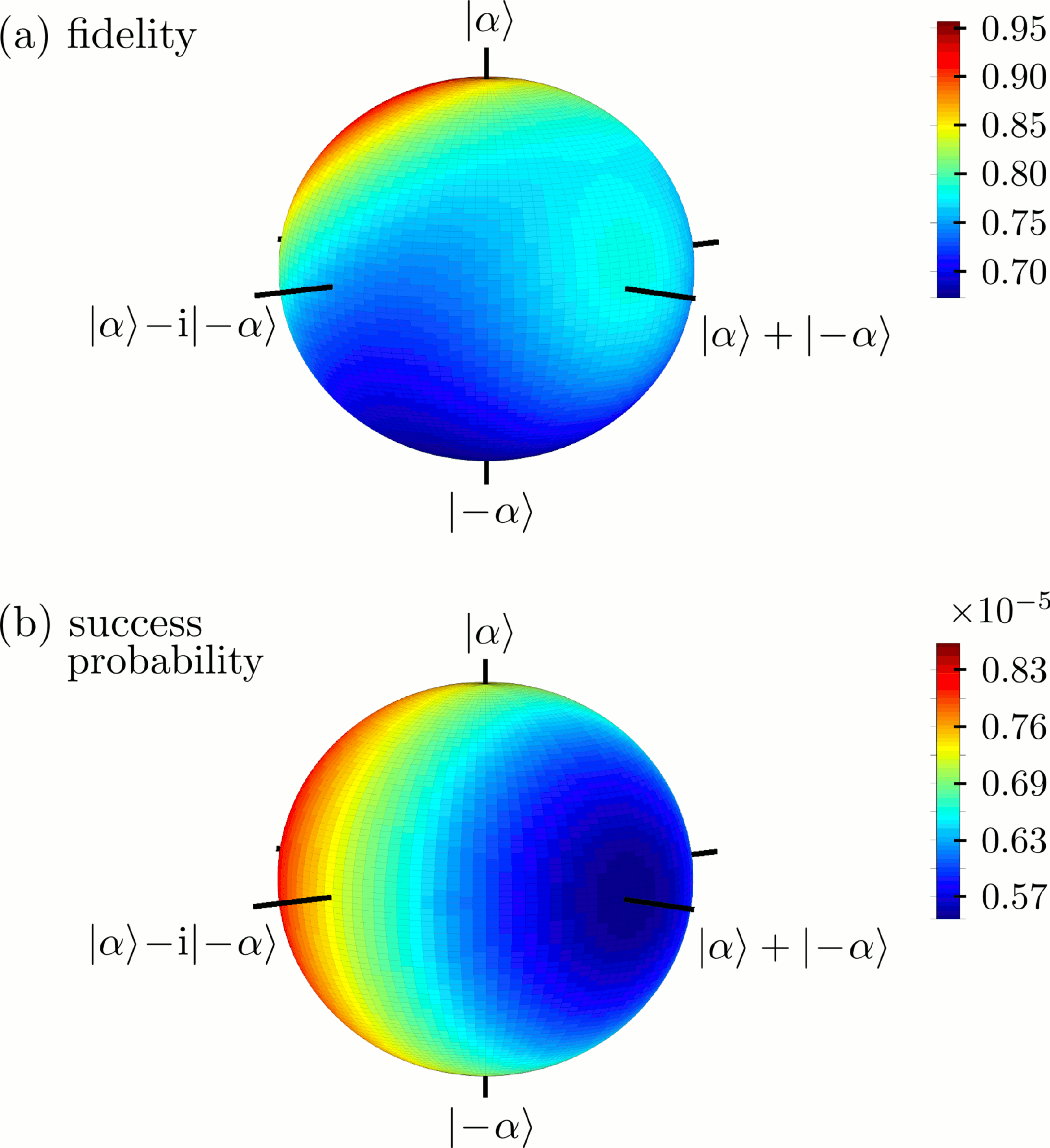}
\caption{The overall quality of the gate is visualized by mapping
the Bloch sphere of input CSQ onto the fidelity $F$ of the output
states (a) and their corresponding success probabilities
$P_{\rm S}$ (b).} 
\label{fig:SimulationRealistic}
\end{figure}

Our simulation starts with an arbitrary qubit in the coherent state
basis, $|\psi_{\rm in}\rangle$, for which the global input state reads
\begin{equation} \label{input_state}
  \hat{\rho}_{\rm in} = |\psi_{\rm in}\rangle_1\langle\psi_{\rm in}| \otimes
  |0\rangle_2\langle 0|\otimes |0\rangle_3\langle 0|\otimes \hat{\rho}^{\rm A}_4,
\end{equation}
where the subscripts are used to label the four participating modes
and $\hat{\rho}^{\rm A}$ represents the density matrix of a squeezed
thermal state used as the ancillary resource. The action of the
gate can now be represented by a unitary evolution of the linear
optical elements, $\hat{U}$, followed by POVM elements of successful
heralding events $\hat{\Pi}$, with the output state given by
\begin{equation} \label{output_state}
  \rho_{\rm out} =
  \frac{1}{P_{\rm S}}\mathrm{Tr}_{123}(
  \hat{U} \hat{\rho}_{\rm in} \hat{U}^{\dag} \hat{\Pi} ),
\end{equation}
where $P_{\rm S} = \mathrm{Tr}( \hat{U} \rho_{\rm in} \hat{U}^{\dag}
\hat{\Pi})$ is the success rate. $\hat{U} = \hat{U}_{23}(t_{\rm BS})
\hat{U}_{12}(t_{\rm ABS1}) \hat{U}_{34}(t_{\rm ABS2})$ is composed
of unitary beam splitter operations coupling the respective modes,
and $\hat{\Pi} = \hat{\Pi}^{\rm HD}_1 \otimes \hat{\Pi}^{\rm APD}_3$
describes the inefficient homodyne and APD measurements.
To parametrize a Bloch sphere of input CSQ states we denote
$u=\cos{\theta}$ and $v=\sin{\theta}\exp({\rm i}\phi)$,
where $\theta\in[0,\pi/2]$ and $\phi\in[0,2\pi]$.
The north and south poles correspond to the pseudo-orthogonal
states $|\alpha\rangle$ and $|\!-\!\alpha\rangle$, respectively.
A mapping of this Bloch sphere onto the corresponding fidelities
and success probabilities at the output is shown in
Fig.~\ref{fig:SimulationRealistic}. The fidelity spans the interval
of $F\in[0.67,0.96]$ with the average value of $\bar{F}=0.78$.
Particularly, for coherent states $|\alpha\rangle$ and
$|\!-\!\alpha\rangle$ at the input, the fidelities of $0.88$ and
$0.67$ are predicted, respectively, which agrees well with the
actually measured values. The success probabilities associated with 
$|\alpha\rangle$ and $|\!-\!\alpha\rangle$ are almost equal which
confirms the correct value of the amplitude quadrature used at
the HD for conditioning. The average success probability is
$\bar{P_{\rm S}}=7.2\times10^{-6}$.

Alternatively, we quantify the performance of the gate by employing
the process fidelity. This quantity is based on the elegant notion
that any operation can be implemented through teleportation:
The desired operation is conducted onto an entangled state which
is subsequently used to teleport the state on which the operation
should be imparted \cite{teleport}.
The quality of such an operation is given by the quality of the
actually transformed entangled state, which can be quantified by
the fidelity with respect to the ideally transformed entangled state.
We have performed a numerical simulation of the transformation of the 
entangled state $|\alpha,\alpha\rangle + |\!-\!\alpha,-\alpha\rangle$
and compared it to the ideally transformed state,
$|\alpha\rangle (|\alpha\rangle +|\!-\!\alpha\rangle)/\sqrt{N_{+}}
+ |\!-\!\alpha\rangle (|\alpha\rangle -|\!-\!\alpha\rangle)/\sqrt{N_{-}}$.
The process fidelity resulting from this simulation reaches ${\cal F}=0.70$.
%{\color{blue}Could we shorten this paragraph slightly if we exceed the page limit? To me it feels a bit injected and not in line with the rest of the paper. I know the referees have not made specific comments about this part. I am referring to the paragraph on the process fidelity!!!}

In conclusion, we have demonstrated a single mode Hadamard gate
for coherent state qubits by using a hybrid projector system
consisting of a conditional homodyne detector and a photon
counter. Its performance has been characterized by a set of basis states and from this we derived a model which could be used to simulate its performance for an arbitrary qubit, following the approach in \cite{Understanding_Photonic_Gates}. This implementation constitutes an important step towards the demonstration of quantum computing with macroscopic
qubit states. To implement universal quantum computing,
the Hadamard gate must be supplemented with a single mode
phase gate (a simplified version was recently implemented
\cite{PIphase_Grangier})
and a two-mode controlled phase gate. In addition
to the implementation of these gates, another outlook is to
refine the experimental techniques or propose new schemes
that may increase the gate fidelity, and thus eventually may
allow for fault-tolerant operation.

\begin{acknowledgments}
The work was financed by the Danish Research Agency
(Project No.~FNU 09-072623) and EU project COMPAS.
PM acknowledges the support by Projects No.~ME10156
of the Czech Ministry of Education and No.~P205/10/P319
of the Czech Grant Agency. MJ acknowledges the support by Projects
No.~MSM6198959213 and No.~LC06007 of the Czech Ministry of Education
and by the Palack{\'{y}} University (PrF\_2011\_015).
\end{acknowledgments}


\begin{thebibliography}{99}
\bibitem{note_1}
The Hadamard transform is the unitary operation defined as
$\hat{H}(x|0\rangle+y|1\rangle)\rightarrow x(|0\rangle+|1\rangle)+
y(|0\rangle-|1\rangle)$, assuming $\langle0|1\rangle=0$. In our case
the coherent basis states $|\alpha\rangle$ and $|\!-\!\alpha\rangle$
are only approximately orthogonal which opens a question of
the normalization of the corresponding CSQ states. Our choice of
the normalized transformations (in contrast to $|\pm \alpha\rangle
\rightarrow |\alpha\rangle \pm |-\alpha\rangle$) has the benefit of
all states being physically sound and, consequently, fidelities
for $\alpha \rightarrow 0$ approaching unity. On the other hand,
it is true that for these small amplitudes the gate does not behave
exactly as the Hadamard gate. However, this is a general problem
caused by the non-orthogonality of the coherent states and we are
avoiding it by choosing reasonably large $\alpha$.
Furthermore, our gate implements both the transformations
(the normalized as well as the unnormalized one), switching between
them just by a suitable choice of the quadrature value used at the HD
for heralding of the gate.
\end{thebibliography}
\end{document}